\documentclass{iitpressproc}

\usepackage{graphicx}
\usepackage{amsmath}
\usepackage{xspace}

\def\pT{\ensuremath{p_{\rm T}}\xspace}
\def\kT{\ensuremath{k_{\rm T}}\xspace}
\def\mT{\ensuremath{m_{\rm T}}\xspace}
\def\sNN{\ensuremath{\sqrt{s_{NN}}}\xspace}

\begin{document}

\title{Kaon Freeze-out Dynamics in \sNN{}=200 GeV Au+Au Collisions at RHIC}

\author{Michal \v{S}umbera for the STAR Collaboration
\address{Nuclear Physics Institute ASCR, 250 68 \v{R}e\v{z}, Czech Republic}}

\maketitle

\begin{abstract}
 Measurements of three-dimensional correlation functions of like-sign low transverse momentum kaon pairs from Au+Au collisions at top RHIC energy \sNN{}=200~GeV are presented. The extracted kaon source function is narrower than the pion one and does not have the long tail along the pair transverse momentum direction. This indicates a much smaller role of long-lived resonance  decays and/or of the emission duration on kaon emission. Three-dimensional Gaussian shape of the kaon source function can be adequately reproduced by Therminator  simulations with resonance contributions taken into account. Comparison to pion data at the same energy reveals that the kaon Gaussian radii in the outward and sideward directions scale with the transverse mass \mT{}. In the longitudinal direction, unlike at lower SPS energies, the Gaussian radii do not seem to follow the exact \mT{} scaling between kaons and pions.
\end{abstract}

\section{Introduction}
The momentum correlations of particles at small relative momenta in their center-of-mass  system contain
an important information about space-time characteristics of the production process on a femtometer 
scale,  so serving as a correlation femtoscopy tool (see, e.g., \cite{led05, led08} and references therein).
For non-interacting identical particles, like photons, these  correlations result from the interference of the production amplitudes due to symmetrization requirement of quantum statistics (QS).
Additional important source of femtoscopic correlations comes from Coulomb and strong final state interaction (FSI).  
It provides an important information on coalescence femtoscopy and correlation femtoscopy with unlike particles,
including the studies of space-time asymmetries in particle production and strong interaction between specific particles \cite{led05}.  

In heavy ion experiments two-particle interferometry with identical charged hadrons has been for a long time used as a reliable technique  to extract the space-time characteristics of the hot expanding fireball \cite{led05, lis05}. Object of study is three-dimensional (3D) correlation function $C(\mathbf{q})$, where $2\mathbf{q}$ is the difference between the momenta of the two particles in the pair center-of-mass system (PCMS).  The correlation function is defined as the ratio of the 3D relative momentum distribution of particle pairs from the same event to pairs constructed from mixed events. In the following a right-handed Cartesian coordinate system with $z$ (\textit{long}) parallel to the beam direction, $x$ (\textit{out}) pointing in the direction of the pair total transverse momentum  and $y$ (\textit{side}) perpendicular to $x$ and $z$ will be used.  $C(\mathbf{q})$ is related to the probability $S(\mathbf{r})$ to emitt a pair of particles with a pair separation vector $\mathbf{r}$ in the PCMS via Koonin-Pratt equation~\cite{lis05}: 
\vskip -.8cm
\begin{equation}
  C(\mathbf{q})-1 \equiv R(\mathbf{q}) = \int d\mathbf{r} K(\mathbf{q},\mathbf{r}) S(\mathbf{r}),~~ K(\mathbf{q},\mathbf{r}) \equiv  
  |\phi(\mathbf{q},\mathbf{r})|^{2} -1.
  \label{3-Dkpeqn}
\end{equation}
\vskip -.3cm 
\noindent The wave function of relative motion of two particle $\phi(\mathbf{q},\mathbf{r})$ incorporates both QS and FSI effects.  The bias arising from the frequently used Gaussian assumption on the source shape ~ \cite{led05, lis05} can be avoided if we directly  extract $S(\mathbf{r})$  inverting  Eq. (\ref{3-Dkpeqn}) numerically~\cite{Brown:1997ku, dan05}. No assumption on the source shape is thus needed.

In \cite{chu08}  data on femtoscopic correlations of like-sign charged pions from  20\% most central Au+Au collisions at \sNN{}=200~GeV were analysed exploiting expansion of the source function $S(\mathbf{r})$ into a Cartesian harmonic basis~\cite{dan05,dan06}:
\vskip -.5cm
\begin{equation}
S(\mathbf{r}) =\sum_{l, \alpha_1 \ldots \alpha_l}S^l_{\alpha_1 \ldots \alpha_l}(r) A^l_{\alpha_1 \ldots \alpha_l} (\Omega_\mathbf{r}),
\label{eqn1}
\end{equation}
\vskip -.3cm
\noindent where $l=0,1,2,\ldots$, $\alpha_i=x, y \mbox{ or } z$, $A^l_{\alpha_1 \ldots \alpha_l}(\Omega_\mathbf{q})$ are Cartesian harmonic basis elements  and $\Omega_\mathbf{q}$ is the solid  angle in $\mathbf{q}$ space. Significant non-Gaussian features including a long range tail in \textit{out} and \textit{long} directions in the pion source function were found. Model comparisons were used to study lifetime and emission duration of expanding fireball. Sizeable emission time differences between emitted pions were required to allow models to be successfully matched to these tails. However, an interpretation of pion correlations in terms of pure hydrodynamic evolution is complicated by significant contributions from later stages of the reaction, such as decays of long-lived resonances and anomalous diffusion from rescattering~\cite{Csanad:2007fr}.

A purer probe of the fireball decay could be obtained with kaons which have less contribution from long life-time resonances and suffer less rescattering than pions. Their lower yields, however, make it difficult to carry out a detailed 3D source shape analysis. A one-dimensional kaon source image measurement for the same colliding system as in Ref.~\cite{chu08} was reported by the PHENIX collaboration ~\cite{aki09}. The measurement corresponds to a fairly broad range of the pair transverse momentum $2\kT$,  which makes the interpretation more ambiguous. In particular, information about the transverse expansion of the system contained in the \kT-dependence of the emission radii is lost. The one-dimensional nature of the measurement also less constrains the model predictions than would be available from a 3D measurement. A different aspect of the fireball expansion can be addressed by studying the \kT{}-dependence of the source size. Data at SPS energies~\cite{Afanasiev:2002fv}, as well as at RHIC  ~\cite{aki09} at relatively higher \kT{} values, showed a scaling behaviour between pions and kaons, as expected from perfect hydrodynamics ~\cite{Csanad:2008gt}.

In this contribution we present recent STAR 3D analyses~\cite{Adamczyk:2013wqm} of the shape and \kT{}-dependent size of the kaon source at mid-rapidity using low transverse momentum like-sign kaon pairs produced  in $\sqrt{s_{NN}}$=200 GeV central Au+Au collisions.

\section{Data analysis}
The kaon source shape was analyzed using 4.6 million 0--20\% central events from 2004, and 16 million 0--20\% central events from 2007. The \kT{}--dependent analysis was carried out using 6.6 million 0--30\% central events from 2004. 
The STAR Time Projection Chamber  was used to select charged kaons with rapidity $|y|<0.5$ and transverse momenta 0.1$<$\pT{}$<$1.0 GeV/c. Only pairs with 0.2$<$\kT{}$<$0.36 GeV/$c$ were accepted. In the \kT{}--dependent analysis, kaon pairs were collected in two bins: 0.2$<$\kT{}$<$0.36 GeV/$c$ and 0.36$<$\kT{}$<$0.48 GeV/$c$. 
The 3D correlation function $C(\mathbf{q})$ defined via Eq.~(\ref{3-Dkpeqn}) was constructed as a ratio of the $N_{\rm {same}}(\mathbf{q})$, for $K^+K^+$ and $K^-K^-$ pairs in the same event to $N_{\rm {mixed}}(\mathbf{q})$. $C(\mathbf{q})$ is flat and normalized to unity over 60$<|\mathbf q|<$100~MeV/$c$. For further experimental details see ~\cite{Adamczyk:2013wqm}.

From the measured 3D correlation function correlation moments 
\vskip -.1cm
\begin{equation}
 R^l_{\alpha_1 \ldots \alpha_l}(q) = \frac{(2l+1)!!}{l!}
 \int \frac{d \Omega_\mathbf{q}}{4\pi} A^l_{\alpha_1 \ldots \alpha_l} 
 (\Omega_\mathbf{q}) \, R(\mathbf{q})
 \label{eqn2}
\end{equation}
\vskip -.1cm
\noindent were extracted.  The lowest correlation moment $R^0$ agrees with the direction independent correlation function $R(q)$ within statistical errors. Even moments with $l>4$ were found to be consistent with zero within statistical uncertainty. As expected from symmetry considerations, the same was also found for odd moments.  Therefore in this analysis, the sum in Eq.~(\ref{eqn1}) is truncated at $l=4$ and expressed in terms of independent moments only. Up to order 4, there are 6 independent moments: $R^0$, $R^2_{xx}$, $R^2_{yy}$, $R^4_{xxxx}$, $R^4_{yyyy}$ and $R^4_{xxyy}$. Dependent moments are obtained from independent ones~\cite{dan05,dan06}. 

Fitting the truncated series to the measured 3D correlation function with a 3D Gaussian,
\vskip -.8cm
\begin{eqnarray}
  S^G(r_x,r_y,r_z) = \frac{\lambda}{ {(2\sqrt\pi)}^3 R_x R_y R_z}
  \exp[-(\frac{r_x^2}{4 R_x^2} + \frac{r_y^2}{4 R_y^2} + \frac{r_z^2}{4 R_z^2})],
\label{eqn3}
\end{eqnarray} 
\vskip -.2cm
\noindent  has yielded the independent moments as a function of $q$. In Eq.(\ref{eqn3}) $R_x, R_y$ and $R_z$ are the characteristic radii of the source in the \textit{out}, \textit{side} and \textit{long} directions, and $\lambda$ represents the overall correlation strength. While $\lambda$ may be sensitive to feed-down from long-lived resonance decays, remainder sample contamination or track splitting/merging not removed by purity and track quality cuts, the radii are virtually independent of these effects. Technically, the fit is carried out as a simultaneous fit on the even independent moments up to $l$=4, yielding $\chi^2/ndf$=1.7 in the source shape analysis, $\chi^2/ndf$=1.1 and $\chi^2/ndf$=1.3 in the 0.2$<$\kT{}$<$0.36 GeV/$c$ and 0.36$<$\kT{}$<$0.48 GeV/$c$ bins in the \kT{}-dependent analysis, respectively.  The three Gaussian radii and the amplitude obtained from this fit are listed in Table~\ref{tab:fits}.

\vspace{-.2cm}
\begin{table}[hbt]
\caption{\label{tab:fits} Parameters obtained from the 3D Gaussian source function fits for the different datasets. The first errors are statistical, the second errors are systematic.}
\begin{tabular*}{\columnwidth}{l @{\extracolsep{\fill}} c @{\extracolsep{\fill}}c @{\extracolsep{\fill}} c}
\hline\hline
Year & 2004+2007   & \multicolumn{2}{c}{2004} \\
Centrality  & 0\%--20\% & \multicolumn{2}{c}{0\%--30\%} \\ \cline{3-4}
\kT	[GeV/$c$]   & 0.2--0.36 & 0.2--0.36 & 0.36--0.48 \\
\hline
$R_x$ [fm] & 4.8$\pm$0.1$\pm$0.2 & 4.3$\pm$0.1$\pm$0.4 & 4.5$\pm$0.2$\pm$0.3 \\
$R_y$ [fm] & 4.3$\pm$0.1$\pm$0.1 & 4.0$\pm$0.1$\pm$0.3 & 3.7$\pm$0.1$\pm$0.1 \\
$R_z$ [fm] & 4.7$\pm$0.1$\pm$0.2 & 4.3$\pm$0.2$\pm$0.4 & 3.6$\pm$0.2$\pm$0.3  \\
$\lambda$  & 0.49$\pm$0.02$\pm$0.05 & 0.39$\pm$0.01$\pm$0.09 & 0.27$\pm$0.01$\pm$0.04 \\
\hline\hline
\end{tabular*}
\end{table}

The shape assumption was tested using a double Gaussian trial function and by pushing the fit parameters to the edges of their errors. Other systematic errors were obtained under varying conditions including magnetic field, data collection periods, charge and various sample purity selections. The systematic errors are largely governed by the limited statistics available. 

\section{Results}

\begin{figure}[hbt]
\includegraphics[width=.55\linewidth,  clip = true, trim = 0. 30. 0. 0. mm]{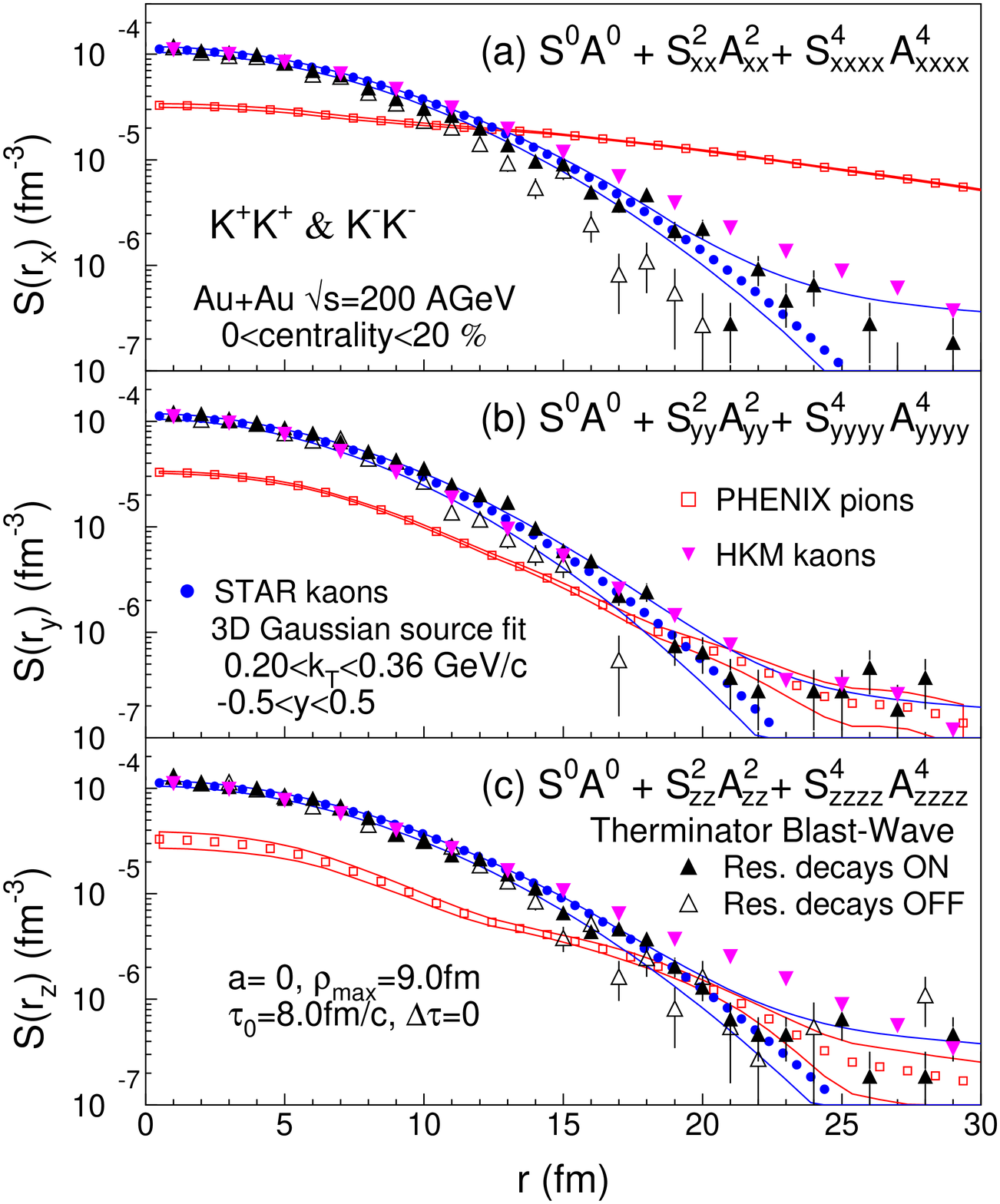}
  \hspace{-.4cm} 
  \vspace{-.8cm}
\includegraphics[width=.52\linewidth, height =.69\linewidth]{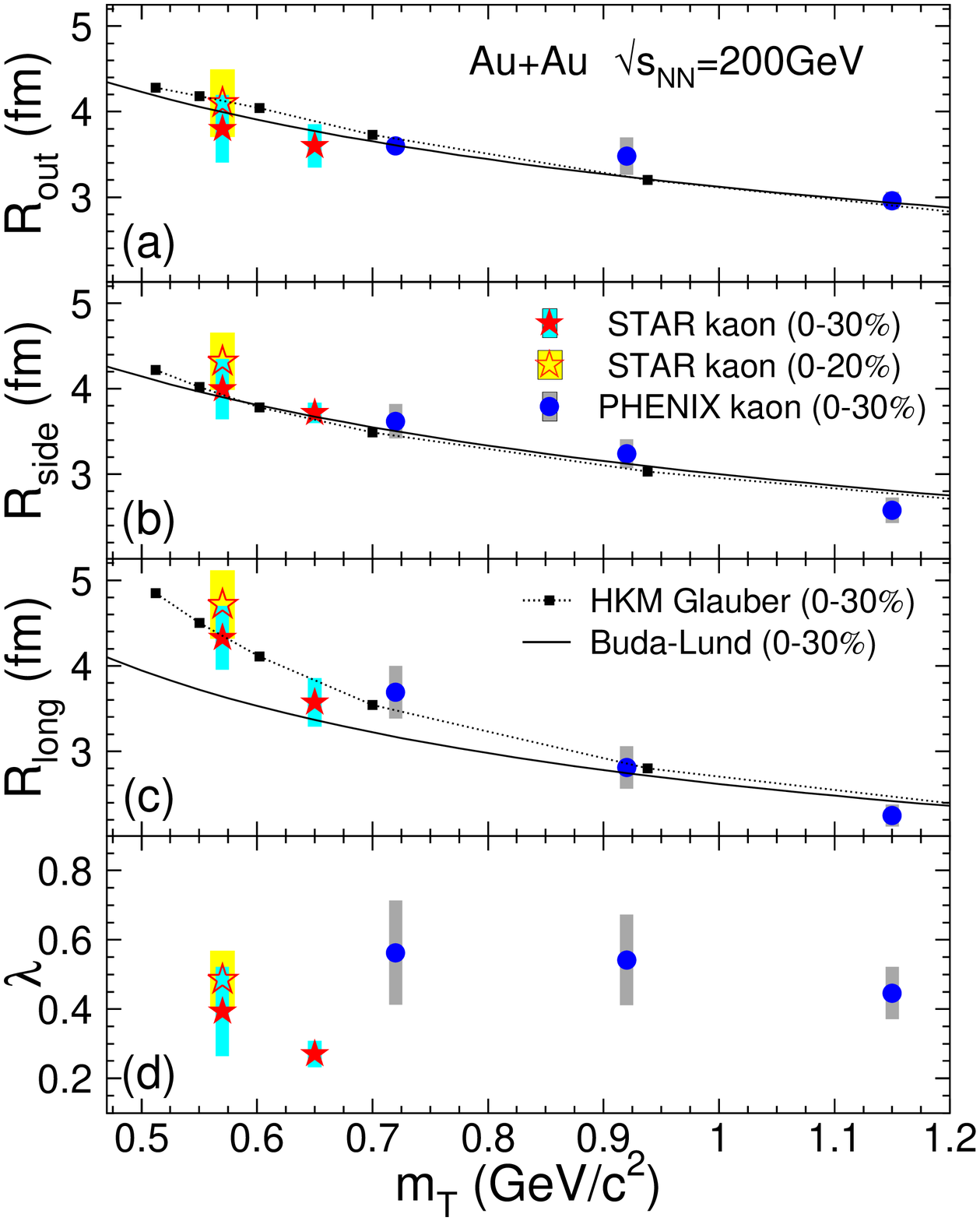}
  \vspace{-.3cm}
  \caption{(Color online) Left: Kaon source function profiles extracted from the data 
    (solid circles) compared to 3D pion source function (squares) from PHENIX
\cite{chu08}, and to the Therminator (triangles pointing upwards) and HKM (triangles pointing downwards) models. Right:
Transverse mass dependence of Gaussian radii and the $\lambda$ for the 30\% most central Au+Au collisions (solid stars). PHENIX data are also plotted (dots). Squares are HKM, solid curves are Buda-Lund model calculations. The 20\% most central data are also shown for comparison (open stars).}
 \label{src_kt}
\end{figure}

The source function profiles in the $x$, $y$ and $z$ directions are shown on the left panel of Figure~\ref{src_kt} (circles). The two solid curves around the Gaussian source function profiles represent the error band arising from the statistical and systematic errors on the 3D Gaussian fit. Note that the latter becomes important for large $r$ values only. The 3D pion source functions from PHENIX~ \cite{chu08} are shown for comparison purposes (squares). While the Gaussian radii are similar, there is a strinking difference between the source shapes of the two particle species, especially in the \textit{out} direction. Note that the PHENIX and STAR pion measurements are fully consistent~\cite{Chung:2010bb}. We have used the STAR tune of the Therminator Blast-Wave model (solid triangles)~\cite{kis05,kis06} to gain a better understanding of this difference. The simulation reproduces the source function profiles with emission duration $\Delta\tau$=0 (solid upward-pointing triangles). However, with resonance contribution switched off, Therminator gives a distribution that is narrower than the measurement (empty triangles). Also note that the pion source function is reproduced with Therminator only when non-zero emission duration is assumed~\cite{chu08}.
Recent simulations of the kaon source function \cite{Shapoval:2013bga} with the hydrokinetic model (HKM)~\cite{kar10} show a good agreement in the \textit{side} direction, although it is slightly over the measurements at larger radii in \textit{out} and \textit{long} (downward-pointing triangles).

\vspace{-1.5cm}

\vspace{-.4cm}
The right panel of Figure~\ref{src_kt}  shows the dependence of the Gaussian radii in the longitudinally co-moving system (LCMS) ($R_{out}$=$R_{x}/\gamma$, ~$R_{side}$=$R_{y}$ and $R_{long}$=$R_{z}$; $\gamma$ is the Lorentz boost in the outward direction from the LCMS to the PCMS frame) as a function of the transverse mass $\mT=(m^{2}+\kT^{2})^{1/2}$. PHENIX kaon data \cite{aki09} are also shown. The perfect fluid hydrodynamics calculations from the Buda-Lund model~\cite{Csanad:2008gt} and the hydrokinetic model (HKM)~\cite{kar10} with Glauber initial conditions are plotted for comparison purposes.
While pions are well described by the Buda Lund model in the whole interval shown~\cite{Csanad:2008gt}, low-\kT kaons in the \textit{long} direction seem to favour HKM over the Buda-Lund model, suggesting that  that contrary to lower SPS energies \cite{Afanasiev:2002fv} the \mT{}-scaling  in the \textit{long} direction is broken at RHIC.

\section{Summary}

We have presented the first model-independent extraction of the 3D kaon source by the STAR Collaboration~\cite{Adamczyk:2013wqm}, at mid-rapidity in $\sqrt{s_{NN}}$=200~GeV central Au+Au collisions, using low-\kT kaon pair correlations and the Cartesian surface-spherical harmonic decomposition technique. No significant non-Gaussian tail has been observed. Comparison with the Therminator model calculations indicates that, although the transverse extent of the source is similar to pions, the shape and the size in the longitudinal direction is very different. This can be attributed to resonance decays, but also indicates that kaons and pions may be subject to different freeze-out dynamics. Although the Gaussian radii follow \mT{}-scaling in the outward and sideward directions, the scaling appears to be broken in the longitudinal direction. Thus the hydro-kinetic predictions~\cite{kar10} are favoured over pure hydrodynamical model calculations. 

\vspace{-.3cm}
\section*{Acknowledgments}
  This research was supported by the grant LA09013 of the Ministry of Education of the Czech Republic.



\end{document}